\documentclass[fleqn,10pt]{wlscirep}
\title{Global-threshold and backbone high-resolution weather radar networks are significantly complementary in a watershed}

\author[1*]{Aurelienne A. S. Jorge}
\author[4]{Iuri da Silva Diniz}
\author[2]{Vander L. S. Freitas}
\author[1]{Izabelly C. Costa}
\author[3]{Leonardo B. L. Santos}

\affil[1]{National Institute for Space Research, Cachoeira Paulista-SP, Brazil}
\affil[2]{Department of Computing, Federal University of Ouro Preto, Ouro Preto, Brazil}
\affil[3]{Center for Monitoring and Early Warning of Natural Disasters (CEMADEN), S\~{a}o Jos\'{e} dos Campos-SP, Brazil}
\affil[4]{Department of Control and Automation Engineering, Federal University of Ouro Preto, Ouro Preto, Brazil}
\affil[*]{aurelienne.jorge@inpe.br}


\begin{abstract}
There are several criteria for building up networks from time series related to different points in geographical space. The most used criterion is the Global-Threshold (GT). Using a weather radar dataset, this paper shows that the Backbone (BB) --- a local-threshold criterion --- generates networks whose geographical configuration is complementary to the GT networks. We compare the results for two well-known similarities measures: the Pearson Correlation (PC) coefficient and the Mutual Information (MI). The extracted backbone network (miBB), whose number of links is the same as the global MI (miGT), has the lowest average shortest path and presents a small-world effect. Regarding the global PC (pcGT) and its corresponding BB network (pcBB), there is a significant linear relationship:  $R2=0.77$ with a slope of $1.15$ (p-value $<E-7$) for the pcGT network, and $R2=0.68$ with a slope of $0.76$ (p-value $<E-7$) for the pcBB network. In relation to the MI ones, only the miGT present a high $R2$ ($0.79$, with slope = $1.95$), whereas the miBB has an $R2$ of only $0.20$ ($\text{slope} =0.24$). On the one hand, the GT networks present a sizeable connected component in the central area, close to the main rivers. On the other hand, the BB networks present a few meaningful connected components surrounding the watershed and dominating cells close to the outlet, with significant statistical differences in the altimetry distribution.
\end{abstract}
\begin{document}

\flushbottom
\maketitle
%
%
\thispagestyle{empty}


\section{Introduction}

Complex networks have been widely applied in the study of several complex systems in nature, and society \cite{barabasi2016network}. In climate, they are used as an alternative tool for investigating climate dynamics \cite{Ferreira_et_al_2021}. Based on long-term events, such studies analyze atmospheric datasets in long time series that range from months to several years \cite{Tsonis2006, Boers2014}.

Boers {\it et al.} (2019) \cite{Boers2019} revealed a global pattern of extreme-rainfall teleconnections using an event synchronization method. They employed a satellite-derived rainfall dataset in a daily temporal resolution for almost 20 years. Tsonis {\it et al.} (2006) \cite{Tsonis2006} identified super-nodes associated with teleconnection patterns based on reanalysis data. Such a dataset comprehended over fifty years in a temporal resolution of one month. Their findings suggest that the organization of teleconnections is related to the stability of the climate system.

Differently, the weather, which deals with short-term changes in the atmosphere, has been little exploited in Network Science. Ceron {\it et al.} (2019) \cite{Ceron2019} published one of the few studies within this context, using spatial and temporal high-resolution data from weather radar to detect community structures. Jorge {\it et al.} (2020) \cite{Jorge2020} also worked with weather radar networks, analyzing the relation between topological and geographical distances.

In this work, we build (geo)graphs (geographical networks \cite{Santos2017}) based on weather radar data, creating connections between points on the geographical space in a watershed. We use different similarity measures: Pearson Correlation (PC) coefficient and the Mutual Information (MI) index \cite{Kraskov_2004}. The former captures linear correspondences between the series, while the latter also recognizes nonlinear relations. Besides, we follow two criteria: a global threshold to connect similar time series and a local threshold criterion to extract the network backbone. Interestingly, both criteria generate significantly complementary network structures. We compare them, taking into account the geographical context. Our findings also show a statistically significant linear relationship between topological and geographical distances.

This paper is organized as follows: Section 2 presents the material and methods, followed by Section 3 that introduces the results and discussion, and the conclusions are in Section 4.

\section{MATERIAL AND METHODS} \label{sec:mm}

\subsection{Weather radar dataset and the Tamanduateí Basin} 

Our study area focuses on the Tamanduateí river's basin in the São Paulo Metropolitan Region, Brazil. Situated on the Tiete river's left margin, the Tamanduatei basin has its source in Mauá. It also crosses the cities of Diadema, São Caetano do Sul, and the eastern and central zones of São Paulo \cite{Ramalho2007}. It is one of the basins with the highest number of extreme rainfall events in the city of São Paulo \cite{COELHO2016}.

We employ the Shuttle Radar Topography Mission (SRTM) Digital Elevation Model data to define the watershed contour. A watershed is a land area that drains rainfalls and streams to a common outlet such as a bay mouth or any point along a stream channel. By employing a digital elevation model, it is possible to identify the drainage system based on the land's topography.

We analyze the precipitation series from January 2015, with data from a weather radar located in the city of São Roque (lat$=23\degree35'56"S$, lon$=47\degree5'52"W$), which is $60$ km distant from our study area. It is an S-band radar with a range of $240$ kilometers in qualitative mode and $400$ kilometers in surveillance mode. This weather radar provides high-resolution data in both space and time domains, with $1$ kilometer and $10$ minutes, respectively. It performs an azimuth scan in $15$ elevation angles, ranging from $0.5$ degrees to approximately $20$ degrees \cite{redemet_2015}. The output is a set of reflectivity values (dBZ units) obtained from the reflectivity factor logarithm (Z), considering the default diameter of rain droplets. Using the Marshall-Palmer formula, reflectivity values correlate to an estimated rainfall rate (R) \cite{Marshall:1947}.

We use only the first azimuth scan and values as they are available in dBZ units. Such data are arranged in a cartesian grid after converted from polar coordinates. From this initial grid, we select only the points strictly inside the limits of the Tamanduateí basin and values that correspond to a precipitation rate above $1$ millimeter per hour.

\subsection{The construction and analysis of the geographical networks}

Spatial embedding is a physical property inherent in many natural phenomena, including the precipitation events we cover in this study. Accordingly, we construct our networks based on a geographical graph, whose nodes have a known geographical location, and their edges are related to an intrinsic spatial dependency \cite{Santos2017}.

The nodes of our graph are the grid points of the weather radar scan, which are inside the Tamanduateí basin area. We preserve its spatial location as an attribute. As a result, if we plot it over a map, we have $587$ nodes, one by square kilometer, inside the basin boundaries. The edges are based on similarities between the corresponding time series of each pair of vertices. We use and compare two similarity functions in this work: the Pearson Correlation (PC) coefficient and the Mutual Information (MI).

The PC is a normalized covariance between the two series, capturing their joint variability. For high PC, the variations in the values of one series also happen in the other. Whether there is a transition from low to high values in one of them, this follows in the other. Linear relations result in a high PC coefficient, whose value is given by

\begin{equation}
\label{eqn:pc}
  r =
  \frac{ \sum_{i=1}^{n}(x_i-\bar{x})(y_i-\bar{y}) }{%
        \sqrt{\sum_{i=1}^{n}(x_i-\bar{x})^2}\sqrt{\sum_{i=1}^{n}(y_i-\bar{y})^2}}.
\end{equation}

 The MI gives a notion of the shared information between the two series. It measures how much knowing one of them reduces the uncertainty related to the other. One could write MI in terms of conditional and joint entropy or joint and marginal probability functions.

\begin{equation}
\label{eqn:mi}
  I(X;Y) =
  \sum_{y\epsilon Y}\sum_{x\epsilon X}p(x,y)log\left(\frac{p(x,y)}{p(x)p(y)}\right)
\end{equation}

\noindent which means that MI recognizes nonlinear relations as well \cite{Shannon_1948} \cite{Kraskov_2004}.

Once all edges have associated similarity measures, one must define which must remain in the network. We use and compare two criteria for that purpose: a global threshold (GT) to connect similar time series and a local strategy to extract the network backbone (BB). In the first one, the edges remain for those pairs whose similarity exceeds a global threshold, defined as the point of maximum diameter of our network \cite{LimaSantos2019}. 

One first builds a similarity matrix $S$ containing the PC or MI for every pair of time series $i$ and $j$. The corresponding adjacency matrix $A$ reads

\begin{equation}
\label{eqn:gt}
A_{ij} = \Theta(S_{ij} - \tau)
\end{equation}

\noindent in which $\Theta$ is the Heaviside step function, that is $1$ for positive arguments and $0$ otherwise, and $\tau$ is the global threshold (GT). $A_{ij}=1$ when the corresponding similarity $S_{ij}$ between time series $i$ and $j$ exceeds $\tau$. 

The BB criterion consists of using each node's weight fluctuations to select the edges to be preserved instead of using a global threshold. Let $k_i$ and $s_i$ be respectively the degree and strength of node $i$. We hypothesize that weights are randomly distributed over the $k_i$ links and sum up to $s_i$. The probability that an incident link has weight $w_{ij}$ or larger is \cite{menczer_fortunato_davis_2020}

\begin{equation}
    p_{ij} = \left ( 1 – \frac{w_{ij}}{s_i} \right )^{k_i -1}.
\end{equation}
   
\noindent The link $(i,j)$ is preserved if $p_{ij} < \alpha$, the desired significance level. Since each edge has two endpoints, we consider the smaller $p_{ij}$. This approach keeps edges that are incident to highly connected nodes or those that substantially contribute to the node strength. The bigger the $w_{ij}$ and $k_i$ are, the smaller the resulting $p_{ij}$ is.

Combining the two similarity metrics, PC and MI, with the two network-building criteria, GT and BB, we end up with four network structures: pcGT, pcBB, miGT, and miBB. We build the pcGT with a global threshold of $0.86$, which results in a graph with $587$ vertices and $1270$ edges. Then, for the pcBB, we adjust the value of $\alpha$ so that the resulting network has approximately the same number of edges as the pcGT. 

When applying the Mutual Information in our dataset, the range of weights is lower than the obtained with the Pearson Correlation coefficient. The threshold of the maximum diameter is $0.6$, and the miGT has fewer edges ($964$) than the pcGT. To build the miBB, we adopt the same idea as before, adjusting the value of $\alpha$ in order to achieve a resulting network with the same number of edges as the miGT.

We compare the mentioned networks with the Configuration model (CM) \cite{configuration_model}, a null model based on random connections. The Networkx library provides an implementation that generates pseudographs by randomly assigning edges to match a given degree sequence. We generate a set of ten thousand samples of pseudographs based on pcGT's degree sequence and another ten thousand samples based on miGT's. We refer to these sets as pcCM and miCM, respectively.

For all the constructed network models, we measure and analyze the network metrics: average shortest path ($\left \langle l \rangle \right.$), average of the local clustering coefficients ($\left \langle c \rangle \right.$), diameter ($D$), heterogeneity parameter ($\kappa$), number of components (NC), the size of the giant component (GC) and the number of singletons (ST). The first one refers to the average of the shortest paths that link every pair of vertices. The local clustering coefficient indicates how connected a node's neighbors are to each other. The diameter is defined as the longest shortest path of a network. The $\kappa$ is the ratio between the squared degree and the square of the average degree ($\kappa = \left \langle k^2 \right \rangle / \left \langle k \right \rangle^2$). Large $\kappa$ means heavy-tailed degree distributions, while $\kappa \approx 1$ approximates to random networks (Poisson distribution). The NC is the number of isolated groups of nodes (components), GC is the size of the largest component, and ST is the number of components with a single node \cite{barabasi2016network}.

Furthermore, we analyze if each network structure could be statistically classified as a small-world network by comparing
the $\langle c \rangle$ and $\langle l \rangle$ between the original network and an equivalent random one with the Erdős and Rényi model \cite{barabasi2016network} --- the same number of nodes and edges, but not necessarily preserving the degree distribution. The small-world phenomenon implies a short distance between two randomly chosen nodes in a network. Considering a graph with $N$ nodes and $L$ randomly positioned edges, one analytically defines a probability $p$ that two random nodes $i$ and $j$ are connected \cite{Albert2002, barabasi2016network}:

\begin{equation}
p = \frac{2L}{N(N-1)}
\end{equation}

\noindent which yields

\begin{equation}
\left \langle k \right \rangle = p(N-1), \;\;
\left \langle c \right \rangle = p, \;\;
\left \langle l \right \rangle = \frac{\log(N)}{\log(\left \langle k \right \rangle)}.
\end{equation}

Statistically speaking, if the original network has a lower $\langle l \rangle$ and a higher $\langle c \rangle$ than its equivalent random network, one says it has a small-world structure.

\section{RESULTS AND DISCUSSION}

Table \ref{tab:metrics} presents the weight (PC coefficient or MI index) range and topological metrics for each network model. Concerning the CM models, the values are the average numbers since they correspond to several realizations. For the same reason, we could not measure the values of $NC$, $GC$, and $ST$ for them. The last two columns contain the $\langle l \rangle$ and $\langle c \rangle$ metrics for the equivalent random network (Erdős and Rényi model) in each case (same number of vertices and edges, without preserving the same degree distribution).

\begin{table}[h]
    \centering
        \rowcolors{2}{gray!25}{white}
        \resizebox{\textwidth}{!}{
    \begin{tabular}{cccccccccccc} 
        \textbf{Network} & \textbf{L} & \textbf{Weight} & \textbf{$\langle l \rangle$} & \textbf{$\langle c \rangle$} & \textbf{$D$} & \textbf{$\kappa$} & \textbf{NC} & \textbf{GC} & \textbf{ST} & \textbf{$\langle l\textsubscript{rand} \rangle$} & \textbf{$\langle c\textsubscript{rand} \rangle$}\\
        {pcGT} & {$1270$} & {$0.86$-$0.98$} & {$8.93$} & {$0.536$} & {$37$} & {$1.775$} & {$125$} & {$349$} & {$85$} & {$4.35$} & {$0.007$}\\
        {pcBB} & {$1269$} & {$0.27$-$0.95$} & {$4.42$} & {$0.225$} & {$19$} & {$3.263$} & {$237$} & {$161$} & {$218$} & {$4.34$} & {$0.007$} \\
        {pcCM} & {$1270$} & {$0.13$-$0.96$} & {$3.89$} & {$0.017$} & {$8.66$}  & {$1.775$} & {$-$} & {$-$} & {$-$} & {$-$} & {$-$}\\
        {miGT} & {$964$} & {$0.6$-$0.85$} & {$10.38$} & {$0.474$} & {$49$} & {$1.422$} & {$82$} & {$305$} & {$33$} & {$5.36$} & {$0.005$}\\
        {miBB} & {$964$} & {$0.18$-$0.69$} & {$3.88$} & {$0.159$} & {$19$} & {$2.962$} & {$202$} & {$152$} & {$152$} & {$5.36$} & {$0.005$}\\
        {miCM} & {$964$} & {$0.04$-$0.74$} & {$5.06$} & {$0.007$} & {$11.5$} & {$1.422$} & {$-$} & {$-$} & {$-$} & {$-$} & {$-$}\\
    \end{tabular}}
    \caption{Topological properties of each network.}
    \label{tab:metrics}
\end{table}

As Table \ref{tab:metrics} shows, the average shortest path ($\langle l \rangle$) and the diameter ($D$) are higher when using a global threshold. The same occurs with the clustering coefficient ($\langle c \rangle$), which increases as the network preserves similar connections. Differently, the heterogeneity parameter ($ \kappa $) is higher when using the BB criterion, as it keeps not only the most similar edges. It is also possible to notice that we generate less fragmented networks using the GT approach. The giant component ($GC$) of the pcGT and the miGT are over twice the size of pcBB's and miBB's, and the number of connected components ($NC$) and singletons ($ST$) are considerably smaller in both GT networks.

Regarding the CM networks, pcCM and miCM, the values of $\langle c \rangle$ and $D$ are the lowest among all the network models, which is typical behavior of random networks. Their $\kappa$ are precisely the same as its correspondent GT network (pcCM$=$pcGT, miCM$=$miGT) since the Configuration Model preserves the degree distribution.

When it comes to analyzing the small-world property, only the miBB fulfills the requirements. It has a lower $\langle l \rangle$ ($3.88$) and a higher $\langle c \rangle$ ($0.159$) when compared with its equivalent random network ($\langle l\textsubscript{rand} \rangle$ = $5.36$, $\langle c\textsubscript{rand} \rangle$ = $0.005$). The miCM could be another example of an equivalent random network, but in this case, considering the average of ten thousand samples and keeping the degree distribution of the original network. In this scenario, using miCM as a comparison base, we also have $\langle l\textsubscript{rand} \rangle$ and $\langle c\textsubscript{rand} \rangle$ that satisfies the conditions to classify the miBB as being statistically a small-world network.

\begin{figure}[hbt!]
\centering
    \begin{subfigure}{.48\textwidth}
		\includegraphics[width=\columnwidth]{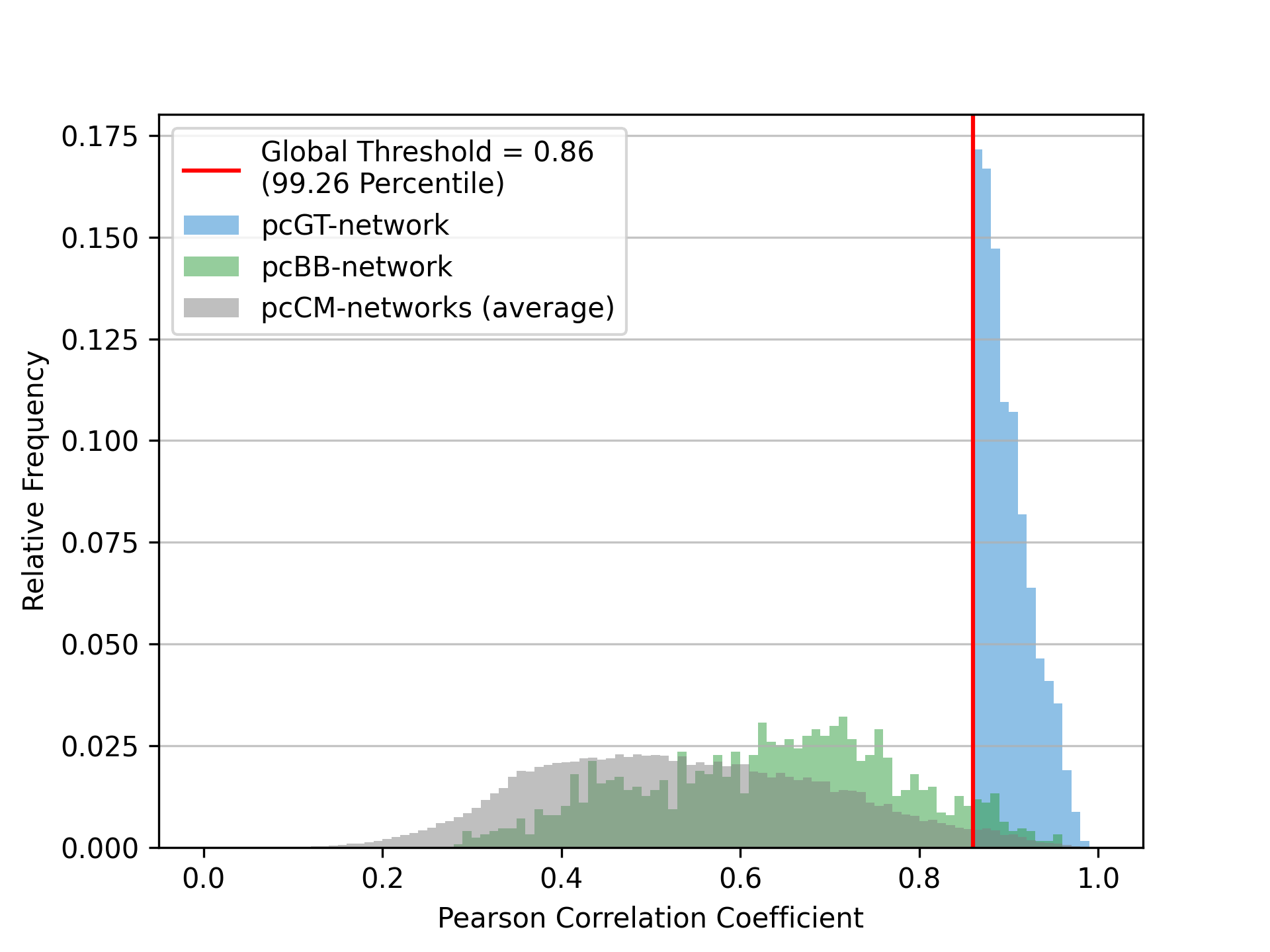}
		\caption{PC-Networks}
		\label{fig:corr_hist_a}
	\end{subfigure}
	\begin{subfigure}{.5\textwidth}
		\includegraphics[width=\columnwidth]{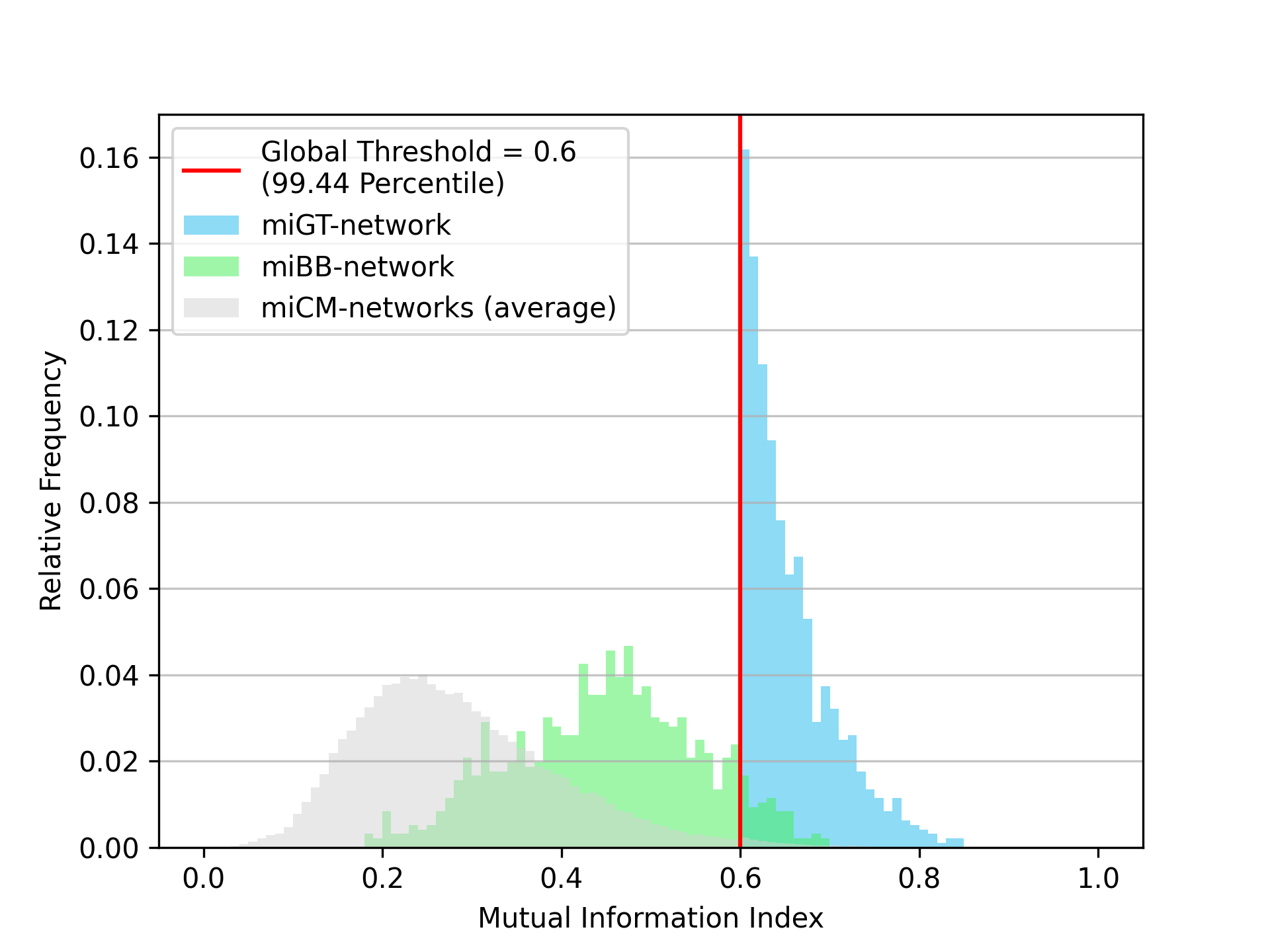}
		\caption{MI-Networks}
		\label{fig:corr_hist_b}
	\end{subfigure}
	\caption{Weight distribution for PC and MI Networks.}
	\label{fig:corr_hist}
\end{figure}

Figure \ref{fig:corr_hist} contains the weight histogram of the investigated networks. Those using Pearson Correlation are on the left (\ref{fig:corr_hist_a}), and the ones with Mutual Information are on the right (\ref{fig:corr_hist_b}). BY DEFINITION, both GT networks (pcGT and miGT) present weights that are higher than or equal to the global threshold. Contrastingly, the pcBB and the miBB networks generate flatter distributions, as the backbone method also considers the nodes' degree and strength besides the edge's weight. The number of shared edges between pcGT and pcBB is $78$ ($6.14 \%$ approximately), and between miGT and miBB is $71$ ($7.36 \%$).

The CM networks present distribution with a higher frequency on lower weights when compared to their corresponding BB or GT. When comparing the set of PC networks with the set of MI, one notices that the shapes of the distributions are similar, with a slight difference in the relative frequencies. MIs have more edges with lower weights, naturally expected since the MI index range is smaller than the PC coefficient range.

\begin{figure}[hbt!]
\centering
    \begin{subfigure}{.48\textwidth}
		\includegraphics[width=\columnwidth]{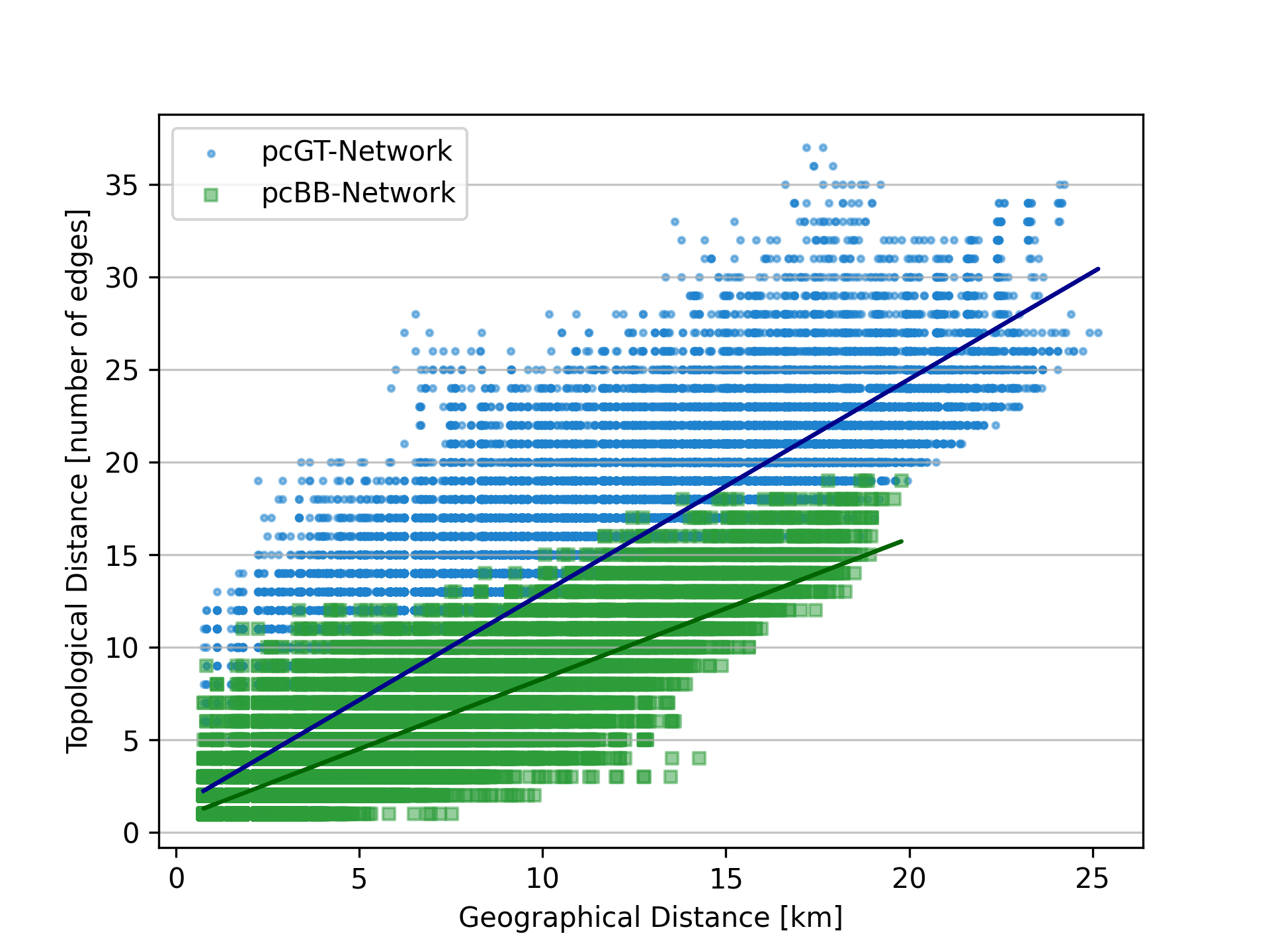}
		\caption{PC-Networks}
		\label{fig:scatter_plot_a}
	\end{subfigure}
	\begin{subfigure}{.48\textwidth}
		\includegraphics[width=\columnwidth]{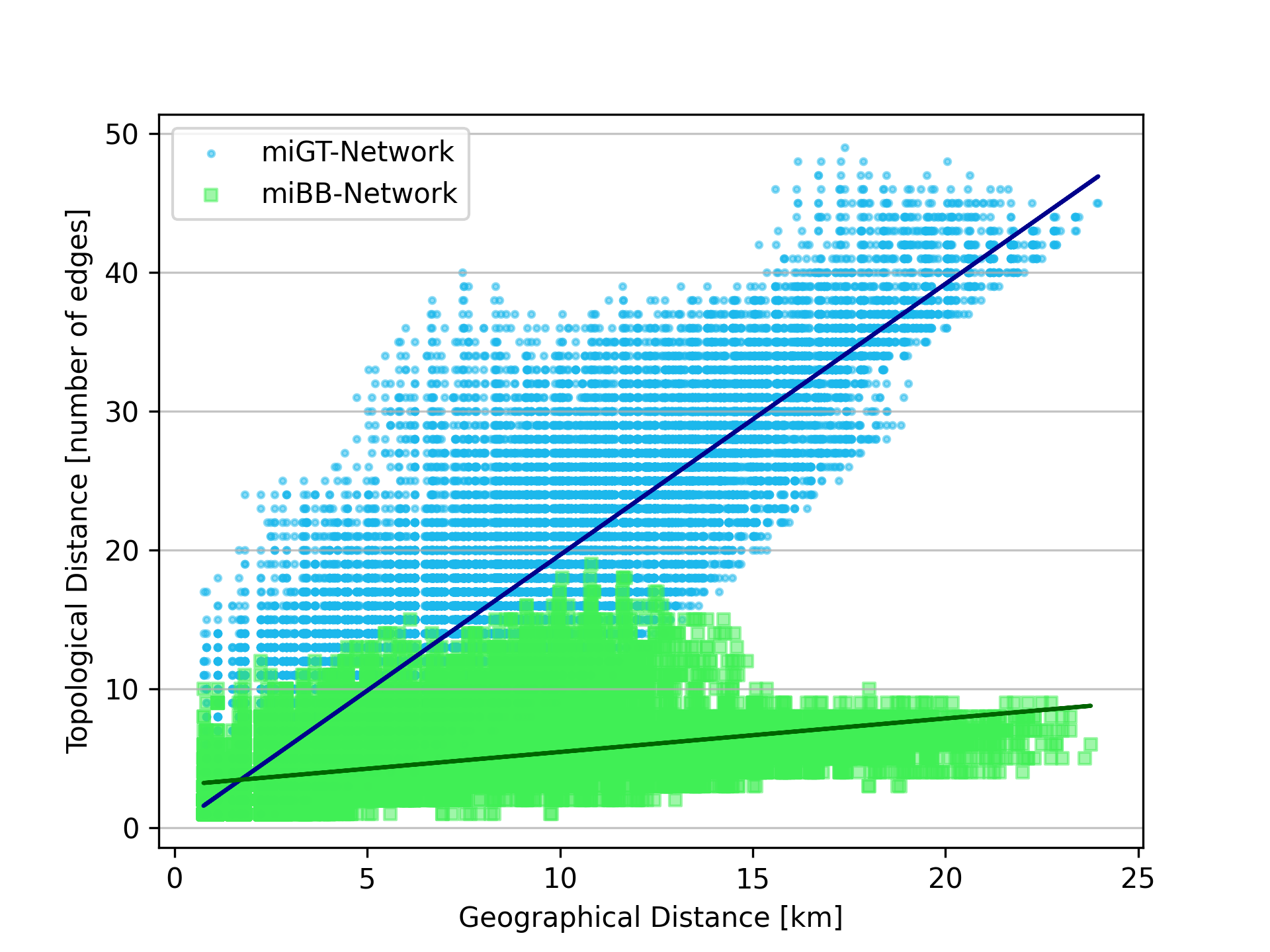}
		\caption{MI-Networks}
		\label{fig:scatter_plot_b}
	\end{subfigure}
	\caption{Topological {\it versus} Geographical distance for each pair of nodes comparing pcGT \textit{versus} pcBB (a), and miGT \textit{versus} miBB (b) --- auxiliar lines representing linear regression, for the GT-networks (blue) and BB-networks (green).}
	\label{fig:scatter_plot}
\end{figure}

Figure \ref{fig:scatter_plot} introduces the scatter plot of the topological distance \textit{versus} the geographical (euclidean) distance - each point represents a pair of nodes. It compares the spatial dependence between GT and BB networks, both based on PC (\ref{fig:scatter_plot_a}) and MI (\ref{fig:scatter_plot_b}). In case of both the PC networks, there is a significant linear relationship:  $R2=0.77$ with a slope of $1.15$ (p-value $<E-7$) for the pcGT, and $R2=0.68$ with a slope of $0.76$ (p-value $<E-7$) for the pcBB. Concerning the MI ones, only miGT presents a high $R2$ ($0.79$, with slope = $1.95$), whereas miBB has an $R2$ of only $0.20$ ($\text{slope}=0.24$).

The pcBB shows longer edges (topological distance = $1$) in geographical space when compared to the pcGT. The pcBB presents an average geographical length of $2.14$, and a maximum geographical length of $7.52$, while pcGT has an average geographical length of $1.06$, and a maximum geographical length equals $2.40$. The pcBB-network contains nodes whose time series are highly correlated to only a few others. However, some highly connected nodes (hubs) are also present due to their combination of high degree and links with lower weights, which allows long-range connections as well (geographical length $> 5$ km).

The behavior is similar when comparing the topological and geographical distances of the MI networks. The BB criterion also presents longer edges (topological distance = $1$), geographically speaking, than the GT criterion. The miBB shows an average geographical length of $2.07$ and a maximum geographical length of $9.78$, whereas the miGT has an average geographical length of $0.94$ and a maximum geographical length of $2.60$. One can notice that the miBB-network has long-distance edges (almost $10$ km) that meaningfully reduce the topological distances inside the network and contribute to its small-world phenomena. Distances of almost $25$ km are reached with less than $10$ edges.

On the other hand, there are nodes in the pcBB and the miBB so close in geographical space ($< 1$km) but relatively far in topological space ($> 5$ edges). This situation is even more apparent for the pcGT and the miGT networks. As BB-networks' connected components are smaller than those from the GT-networks', topological distances are smaller for the BB-networks than for the GT-networks. In terms of geographical space, the miBB presents a greater reach than the pcBB because of its long edges.

\begin{figure}[hbt!]
\centering
    \begin{subfigure}{.49\textwidth}
		\includegraphics[width=\columnwidth]{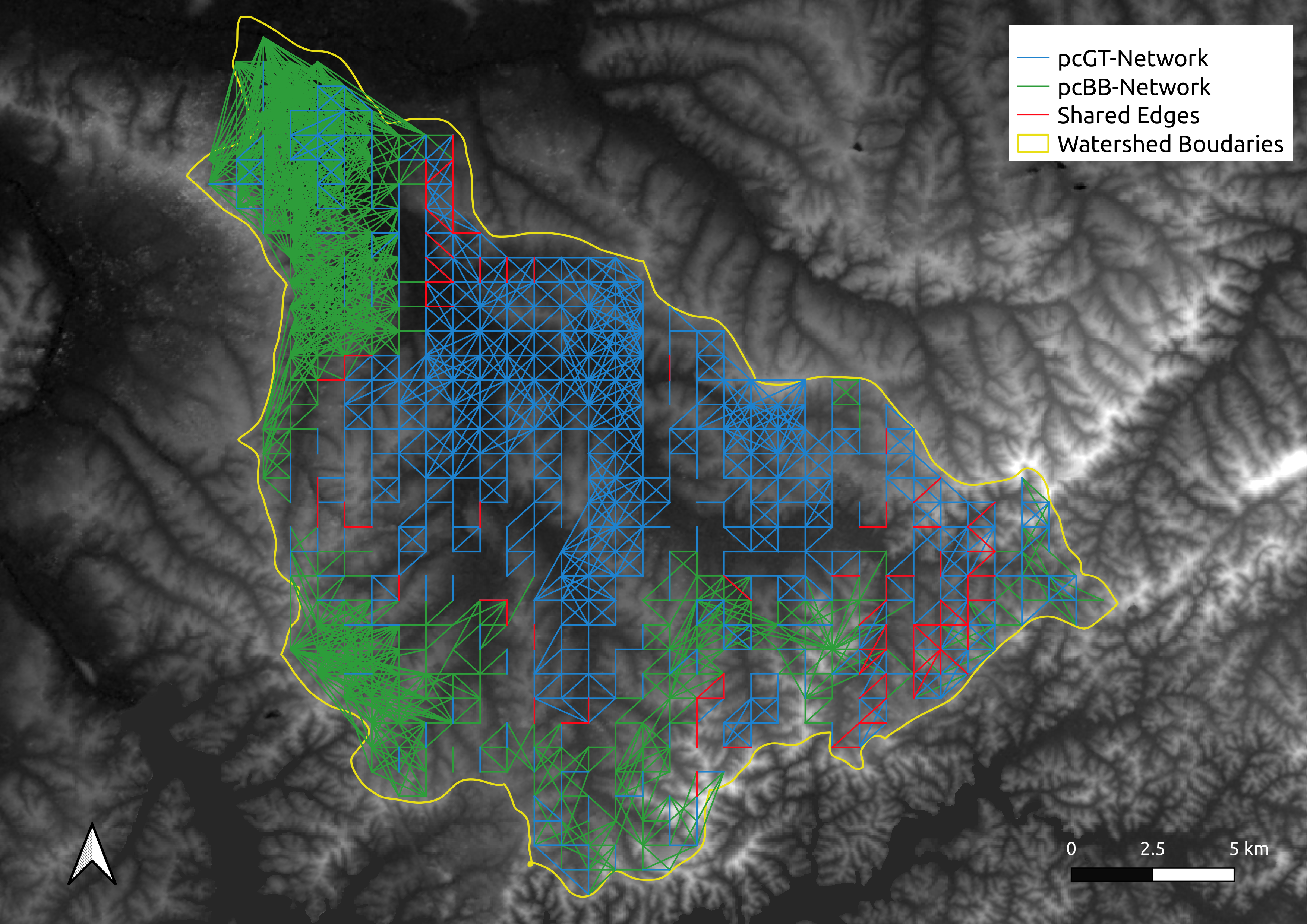}
		\caption{pcGT and pcBB networks}
		\label{fig:map_pc}
	\end{subfigure}
	\begin{subfigure}{.49\textwidth}
		\includegraphics[width=\columnwidth]{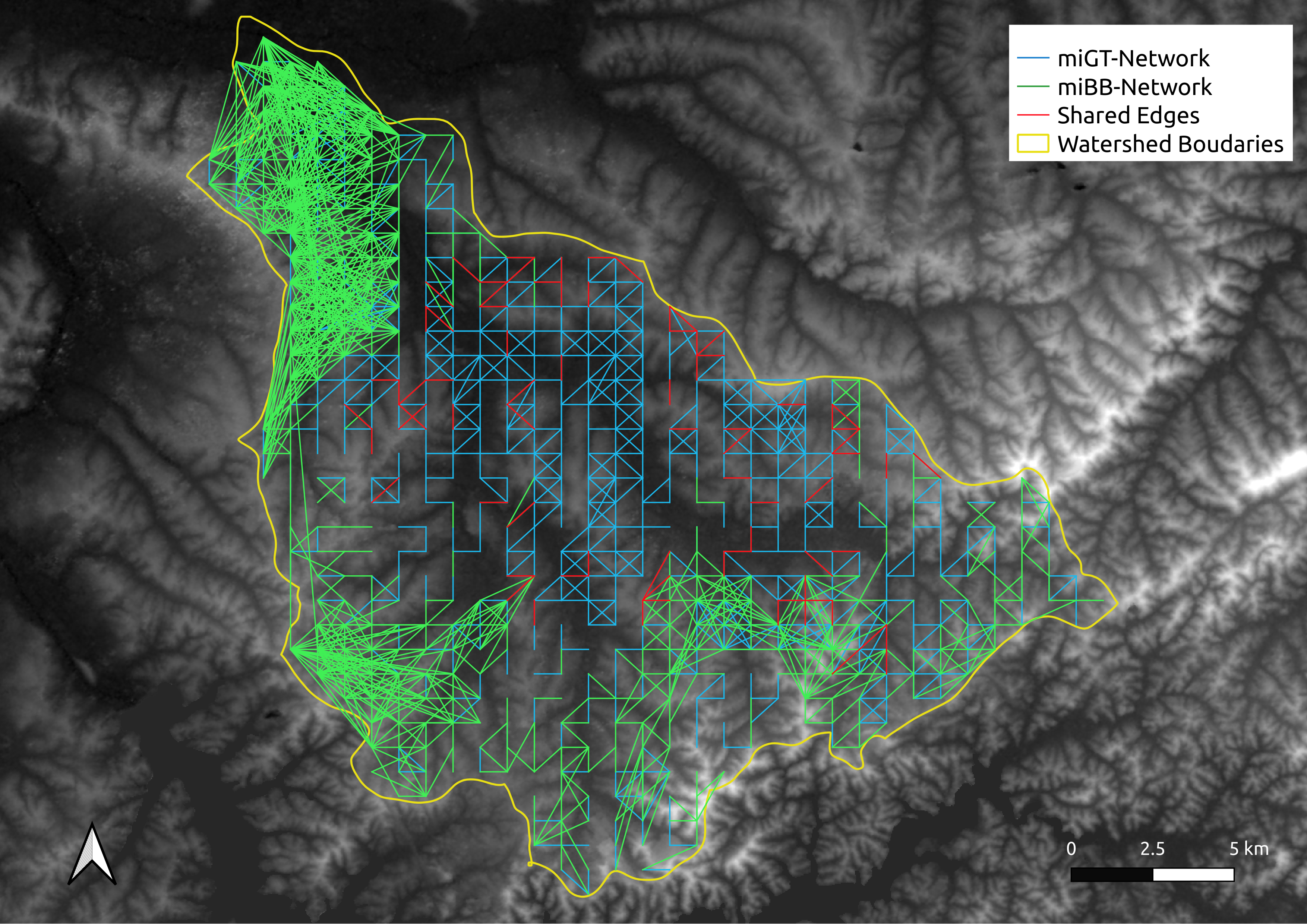}
		\caption{miGT and miBB networks}
		\label{fig:map_mi}
	\end{subfigure}
	\caption{Networks on the watershed (contour in yellow): pcGT and pcBB (a), miGT and miBB (b). GT-network (blue), BB-network (green), and shared edges (in red). In the background, SRTM altimetric data (the lower a cell, the darker it is). }
	\label{fig:maps}
\end{figure}

One can visually verify the spatial structure of the PC and the MI networks in Figure \ref{fig:maps}. The intersection between GT and BB structures, for both similarity metrics, is represented by the edges in red. The GT and the BB networks are significantly complementary in the studied watershed. A watershed represents the set of points on the space with a standard outlet for surface runoff. In the background, SRTM altimetric data is employed and the lower a cell, the darker it is. The structure of PC and MI networks are very similar. The BB networks surround the watershed, mainly in the higher part of it, southwest, and around the outlet. Oppositely, the GT networks are mainly on the central watershed area, connecting cells in a region with no high difference of altimetry and high correlation in rainfall time series. When comparing the PC with the MI networks, there are slight observable differences in the map. The miGT is visually less clustered than the pcGT, confirming the average clustering coefficient values in Table \ref{tab:metrics} (pcGT's $\langle c \rangle = 0.536$, miGT's $\langle c \rangle = 0.474$). The miBB has long edges on the west boundary of the watershed that probably contributes to a lower average shortest path, and consequently, to the small-world effect.

\section{CONCLUSION}

This paper presented different structures of geographical networks based on weather radar data, using two similarity measures: the Pearson Correlation (PC) coefficient and the Mutual Information (MI) index. Furthermore, we employed distinct criteria to create its connections: a Global-Threshold (GT), a Backbone (BB), and a Configuration Model (CM).

The scatter plot of the topological distance \textit{versus} the geographical (euclidean) distance revealed a statistically significant linear relationship for the pcGT and miGT networks and even for the pcBB one. The miBB was an exception presenting a low average shortest path for distant geographical points. Furthermore, our analysis showed that it could be considered a small-world network from a statistical point of view.

It is well known that the GT criterion returns a network linking nodes with very similar behaviors (regarding its time series). On the other hand, the BB criterion provides a network linking nodes with a higher level of heterogeneity depending on the relation between the link's weight and node's strength \cite{menczer_fortunato_davis_2020}. This paper presented a geographical analysis for the different network structures in the context of a watershed: the GT networks are in the central area, close to the main rivers, while the BB networks surround the watershed and dominate cells close to the outlet. Using both PC and MI, the number of shared edges between GT and BB is only around 7\% of the total number of edges, showing a significant complementarity.

As future work, we intend to reproduce the analysis in several other watersheds, from mountainous regions to floodplains, looking for spatial signatures for the different networks in the different landscapes.


\bibliography{sample}


\section*{Acknowledgements}

This Research was partially supported by grant 420338/2018-7 of the Brazilian National Council for Scientific and Technological Development (CNPq) and by grants 2018/06205-7 and 420338/2018-7 of São Paulo Research Foundation (FAPESP) and DFG-IRTG 1740/2.







\end{document}